\documentclass[%
reprint,
amsmath,amssymb,
aps,
pra,
floatfix,
]{revtex4-2}

\usepackage[english]{babel}
\usepackage{amsmath,amssymb,amsfonts}
\usepackage{graphicx}
\usepackage{braket}
\usepackage{algorithm}
\usepackage{algpseudocode}
\usepackage{qcircuit}
\usepackage[caption=false]{subfig}
 
\newcommand{\perasure}{p_{\rm e}}
\newcommand{\pdep}{p_{\textrm{dep}}}
\newcommand{\stabilizergroup}{\mathcal{S}}
\newcommand{\stabilizergenerators}{\mathcal{G}}

\newcommand{\secref}[1]{Sec.~\ref{#1}}
\newcommand{\figref}[1]{Fig.~\ref{#1}}

\begin{document}
\preprint{APS/123-QED}

\title{Erasure-tolerance scheme for the surface codes on neutral atom quantum computers}
\author{Fumiyoshi Kobayashi}%
\email{fkobayashi@qi.mp.es.osaka-u.ac.jp}
\affiliation{Graduate School of Engineering Science, Osaka University,
	1-3 Machikaneyama, Toyonaka, Osaka 560-8531, Japan.}
\affiliation{%
 Center for Quantum Information and Quantum Biology, Institute for Open and Transdisciplinary Research Initiatives, Osaka University, Japan.
}%

\author{Shota Nagayama}%
\affiliation{%
 Mercari R4D, Mercari Inc., Roppongi Hills Mori Tower 18F, 6-10-1, Roppongi, Minato-ku, Tokyo 106-6118, Japan.
}%
\affiliation{%
 Graduate School of Media and Governance, Keio University, 5322 Endo, Fujisawa-shi, Kanagawa 252-0882 Japan.
}%
\date{\today}


\begin{abstract}
Neutral atom arrays manipulated with optical tweezers are promising candidates for fault-tolerant quantum computers due to their advantageous properties, such as scalability, long coherence times, and optical accessibility for communication. A significant challenge to overcome is the presence of non-Pauli errors, specifically erasure errors and leakage errors. Previous work has shown that leakage errors can be converted into erasure errors; however, these (converted) erasure errors continuously occur and accumulate over time. Prior proposals have involved transporting atoms directly from a reservoir area—where spare atoms are stored—to the computational area—where computation and error correction are performed—to correct atom loss. 
While coherent transport is promising, it may not address all challenges—particularly its effectiveness in dense arrays and the efficacy of alternative methods should be investigated.
In this study, we evaluate the effects of erasure errors on the surface code using circuit-based Monte Carlo simulations that incorporate depolarizing and accumulated erasure errors. We propose a new scheme to mitigate this problem: a \textit{$k$-shift erasure recovery} scheme. Our scheme employs code deformation to repeatedly transfer the logical qubit from an imperfect array with accumulated erased qubits to a perfect array, thereby tolerating many accumulated erasures. Furthermore, our scheme corrects erasure errors in the atom arrays while the logical qubits are evacuated from the area being corrected; thus, manipulating optical tweezers for erasure correction does not disturb the qubits that constitute the logical data. Our scheme provides a practical pathway for neutral atom quantum computers to achieve feasible fault tolerance.
\end{abstract}


\maketitle
\section{Introduction}\label{sec:intro}

Scalable and universal quantum computers are expected to be the next generation beyond classical computers. Among the various platforms, neutral atom quantum computers are particularly promising candidates, as they have demonstrated key functionalities required for fault-tolerant quantum computation~\cite{Bluvstein2023-sm}. Rydberg atoms possess desirable properties as qubits, such as long coherence times~\cite{Young2020-st, Jenkins2022-im, Klusener2024-yu, Ma2022-dt} and high controllability in arrays~\cite{Barredo2016-sb, Barredo2018-ts, Bluvstein2022-ll, Ma2022-dt}. 

However, significant hurdles remain to be overcome, including atom loss—the disappearance of qubits caused by the escape of atoms from optical tweezers—and leakage errors, which occur when atom states leak out of the computational qubit space due to imperfect operations during the execution of quantum gates.
To address those problems, it has been shown that leakage errors and atom loss are detectable using Rydberg gates~\cite{gottesman1997stabilizer, Preskill1997-ii, Saffman2016-br, Chow2024-fb}, and methods to convert leakage errors induced by Rydberg excitation to erasure errors have also been proposed~\cite{Wu2022-uy, Ma2023-jp, Scholl2023-gw}. Therefore, these techniques allow atom loss and leakage errors to be treated as erasure errors.

Fortunately, surface codes are known to tolerate erasure errors using super stabilizers~\cite{Stace2010-ka, Nagayama2017-gq}. However, these studies have not revealed the limitations of the surface code when erasure errors increase dynamically and accumulate on an array—a behavior commonly observed in neutral atom quantum devices.

A challenge in achieving erasure tolerance with the surface code is refilling vacant spots with reservoir atoms. Previous studies have proposed refilling erased atoms directly into the online array on which an error-correcting code is encoded~\cite{Cong2021-bj, Wu2022-uy}. 
Recently, atom transport approaches have been demonstrated, wherein atoms are coherently transported without disturbing surrounding qubits on sparse arrays where the atoms are far apart enough to allow one to move between them~\cite{Bluvstein2022-ll, Bluvstein2023-sm}.
While coherent transport is promising, it may not address all challenges—particularly its effectiveness in dense arrays, which can integrate more qubits within a given space than sparse arrays. 
Transporting atoms may have side effects that cause atom loss when the transported atoms are close to other atoms, e.g., closer than $5$~$\mu \mathrm{m}$~\cite{Barredo2016-sb}. 
This limitation makes the development of alternative methods worthwhile. 
These methods may complement each other in practical quantum computer architectures by supporting different array densities, operating under varying erasure or physical error rates, and requiring different classical control software.


In this study, we evaluate the performance of the surface code under depolarizing errors and dynamically accumulating erasure errors. We assume that erasure errors are not corrected immediately and accumulate because the array is too dense to transport atoms between sites. Although increasing the code distance does not suppress the effect of our erasure error model—since the theoretical threshold is zero in the asymptotic limit of the code distance—we are particularly interested in the behavior at finite code distances toward feasible implementations of the quantum error correction. To investigate this, we performed circuit-based Monte Carlo simulations to observe how the surface code performs under finite code distances in the presence of depolarizing errors and dynamically accumulating erasure errors.

Initially, we simulate a surface code of distance $d$ with $d$ rounds of measurements, which is the minimum required for the syndrome-measurement circuit with measurement errors. The simulation shows that the surface code does not exhibit a uniform and explicit threshold for the depolarizing error rate and only demonstrates a pseudo-threshold that holds only for relatively short code distances for the depolarizing error rate under dynamically accumulating erasures. In scenarios beyond a certain erasure rate, the surface code at any code distance does not function effectively as an error-correcting code, and no threshold of depolarizing error rate exists. This indicates that accumulated erasure errors can destroy the code. Conversely, in scenarios with a lower erasure error rate, the surface codes function as error-correcting codes and suppress the logical failure rate by increasing the code distance. Consequently, threshold-like behavior can be observed. These results suggest that increasing the code distance at higher erasure error rates makes the code more vulnerable to failure, as the occurrence of erasure errors rises with the number of rounds of syndrome measurements.

As a consequence of our simulations, we find that the number of accumulated erasure errors directly contributes to the destruction of the code. Even if a sufficiently low Pauli error rate is maintained for a given code distance, too many accumulated erasure errors will destroy the code if the number of erasure errors cannot be reduced.

To address this problem, we propose a new scheme to tolerate erasure errors, namely the \textit{$k$-shift erasure recovery}. Our scheme protects the logical state from erasure errors by combining the code deformation technique of the surface code with atom transportation to rearrange or refill atoms in the array. This scheme maintains the array without concern for the atom loss caused by atom transportation via tweezers, as it separates the operation of maintaining the logical qubit from the maintenance of the atom array. 
The numerical simulation shows the $k$-shift erasure recovery improves the logical failure rate compared with the simple repetition of syndrome measurements on a single array.

Our scheme advances the development of Rydberg quantum computers into fault-tolerant quantum computers by protecting a logical quantum state from both Pauli errors and erasure errors.

\section{Rydberg Atom Qubits in Optical Tweezer Arrays and Their Error Model}

Neutral atom quantum computers commonly use atom species such as alkali metals (e.g., Rb and Cs) and alkaline earth metals (e.g., Yb and Sr). These atoms are referred to as Rydberg atoms when their interactions are controlled by exciting them to Rydberg states—states with a large principal quantum number. Multi-qubit gates among neutral atoms can be implemented using these interactions, specifically through the Rydberg blockade mechanism~\cite{Jaksch2000-fh, Saffman2016-br}.


Optical tweezer arrays are arrays of tightly focused laser spots, generated by sending laser beams into a high-NA objective lens. Two common techniques used to generate such arrays are spatial light modulators (SLMs) and acousto-optic deflectors (AODs)~\cite{Saffman2016-br}. SLMs can create arbitrarily-shaped static tweezer arrays, while AODs specialize in moving atoms on a microsecond timescale. This combination is highly favorable for realizing large, reconfigurable atom arrays~\cite{Endres2016-ez, Barredo2018-ts, Bluvstein2022-ll}.


A single atom has multiple energy levels; two of these levels can be chosen to define a qubit, such as levels within a hyperfine structure. Arbitrary single-qubit gates are realized using Rabi oscillations between these qubit sublevels~\cite{Saffman2016-br}. Multi-qubit gates are achieved via van der Waals interactions between atoms in Rydberg states, known as Rydberg gates.

Let $r$ be the distance between atoms. The dipole-dipole interaction decreases proportionally to $1/r^6$. Rydberg gates can be applied between atoms that are not necessarily nearest neighbors in the array, provided they are close enough for the dipole interaction to act effectively~\cite{Radnaev2024-lw,Pecorari2025-bl}. The typical two-qubit Rydberg gate is the controlled-Z (CZ) gate~\cite{Jaksch2000-fh}. A controlled-NOT (CNOT) gate can be constructed by surrounding a CZ gate with Hadamard gates on the target qubit. By combining arbitrary single-qubit gates and multi-qubit gates, a universal quantum gate set is achieved on neutral atom quantum computers.


Neutral atom qubits on the tweezer arrays are subject not only to Pauli errors within the qubit space but also to leakage errors, where the state leaks to levels outside of the qubit space, and atom loss, where atoms disappear from optical tweezer sites. Some leakage states can be returned to the qubit space through optical pumping~\cite{Cong2021-bj}, while other leakage errors can be converted to erasure errors if their occurrence is detected~\cite{Wu2022-uy}. Alkaline-earth atoms have demonstrated the erasure conversion protocol for leaked states due to Raman scattering from the tweezer light within the qubit manifold~\cite{Ma2023-jp} and imperfect state preparation~\cite{Scholl2023-gw}.

Atom loss refers to the disappearance of atoms from their tweezer sites. The presence or absence of an atom at a site can be detected using quantum circuits
which employs an ancilla qubit~\cite{gottesman1997stabilizer, Preskill1997-ii, Saffman2016-br}. 
The cesium atom array has demonstrated
the detection of atom loss via this circuit~\cite{Chow2024-fb}.
Thus, atom loss can also be treated as erasure errors.

By combining these features, non-Pauli errors, including leakage errors and atom loss, can be treated as erasure errors within the software domain of error correction, provided their occurrences are detectable. An erasure error maps a quantum state $\rho$ to a non-qubit space with probability $\perasure$, i.e.,
\begin{equation}\label{eq:erasure_error}
\mathcal{E}_e (\rho) = (1-\perasure)\rho + \perasure\ket{e}\bra{e},
\end{equation}
where $\ket{e}$ is a state outside the qubit space~\cite{Grassl1996-ua}.

The erasure correction for these two types of erasure errors are assumed to be achieved by atom replacement in this work.
For the atom loss, we refer to the process of reintroducing an atom into a tweezer site where it has been physically atom lost as atom replacement.
For the leakage error, we assume that the state cannot be returned to the computational basis when a leakage error is detected. 
In such cases we physically remove atoms under leakage error and correct the vacant tweezer sites using atom replacement.
Therefore, the both erasure corrections are achieved by atom replacement.

\section{Surface Code with Erasure Errors}\label{sec:surfacecode}

\subsection{Surface Code}

\begin{figure}[htbp]
\centering
\subfloat[Super stabilizer\label{fig:SurfaceCode}]{\includegraphics[width=0.4\linewidth]{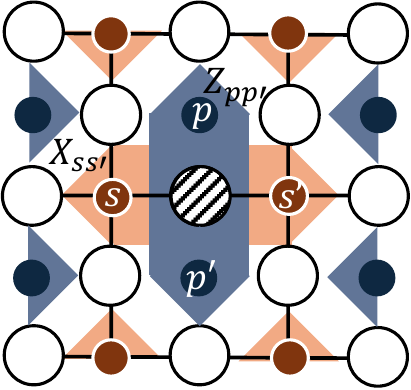}}\
\subfloat[Time evolution of stabilizers with accumulated erasure errors \label{fig:time-dependent-erasure}]{\includegraphics[width=0.85\linewidth]{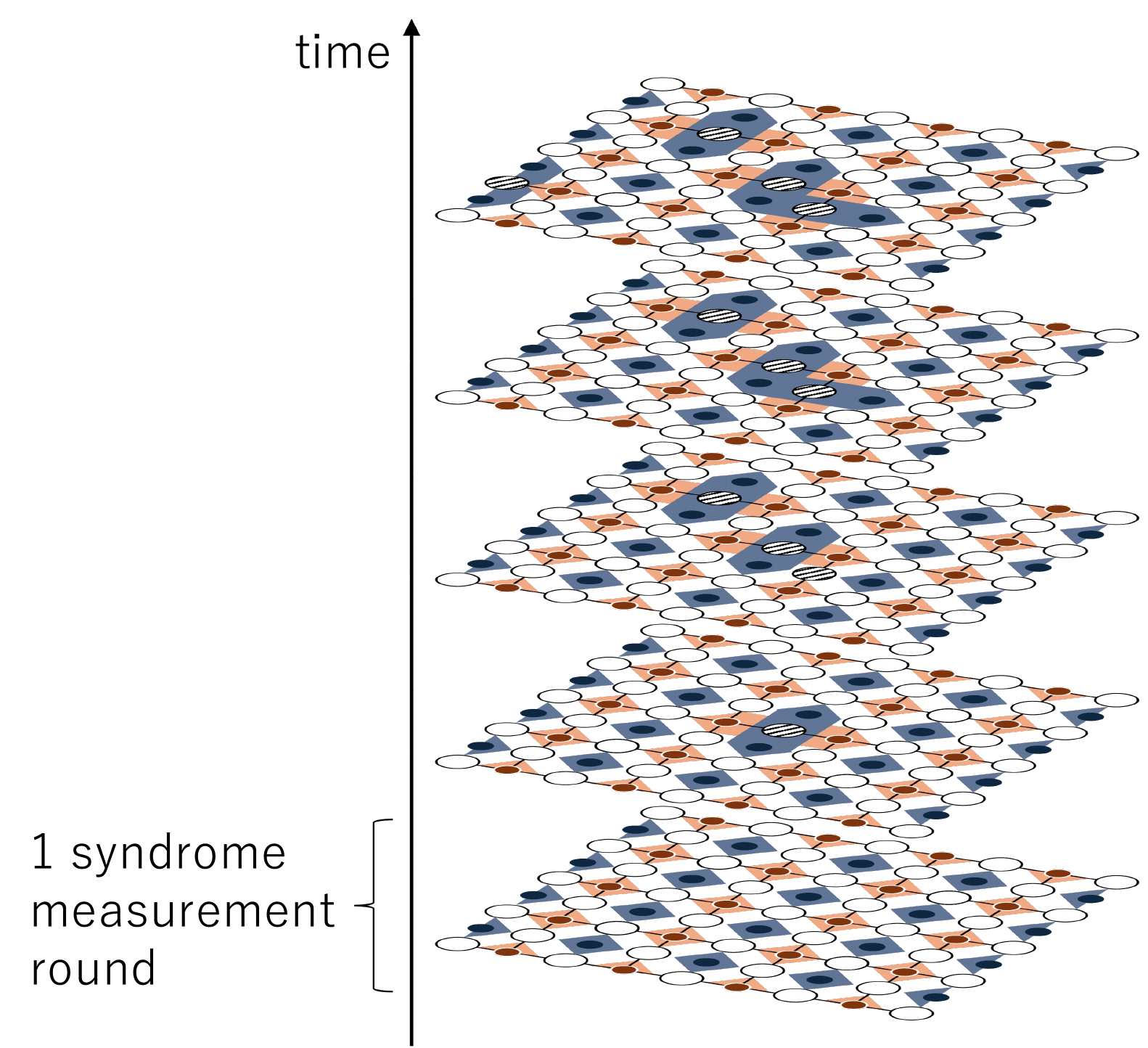}}\
\centering
\subfloat[Syndrome measurement circuit for the surface code \label{fig:syndrome_circuit}]{\includegraphics[width=0.9\linewidth]{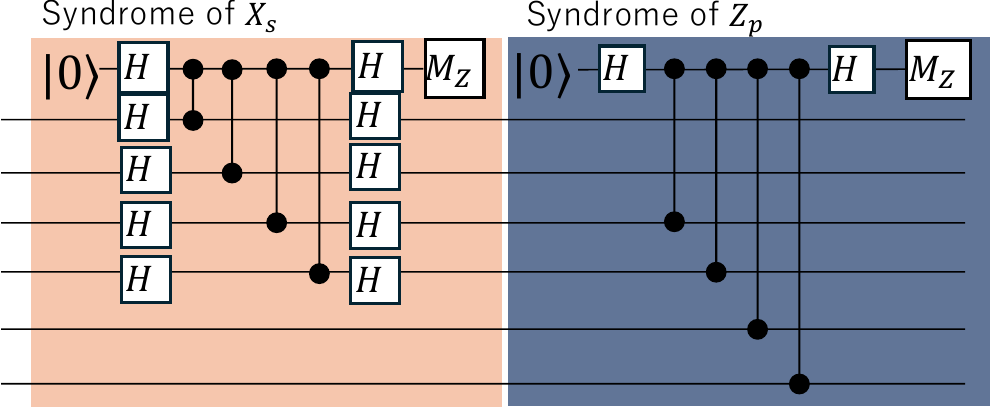}}\
\caption{(a)~Illustration of the stabilizer generators $\mathcal{\tilde{G}}$ of the surface code. The large white circles represent data qubits. The orange (blue) squares represent stabilizer generators of $XXXX$ ($ZZZZ$) for star (plaquette) operators. The small orange (blue) circles are ancilla qubits for syndrome measurements. Suppose a data qubit indicated by a circle with diagonal lines disappears; in that case, the stabilizer generators around it are merged. Then, a super star (plaquette) operator of a large orange (blue) polygon is added as a new stabilizer generator to maintain the functionality as an error-correcting code. 
(b)~An example of the time evolution of a qubit array encoded by the surface code with distance $5$. 
The bottom layer illustrates the situation before the syndrome measurement, with no erased qubits. The vertical axis represents time, showing how the erased qubits (circles with diagonal lines) increase as time progresses. 
(c)~A circuit diagram of syndrome measurements for the surface code. The shaded regions in blue and orange represent syndrome measurements of $X^{\otimes 4}$ and $Z^{\otimes 4}$, respectively.}
\end{figure}

The surface code is a type of topological quantum error-correcting code~\cite{Dennis2001-jj, Horsman2011-yv}. Consider a qubit array arranged on a square lattice, where the data qubits (white circles) are located on each edge, and the ancilla qubits for error syndrome measurements (blue and orange circles) are placed on the faces and vertices, as shown in \figref{fig:SurfaceCode}. To define the stabilizer group $\stabilizergroup$ of the surface code for an ideal array without erasures, we introduce the star and plaquette operators:

\begin{equation}
X_s = \bigotimes_{e\in\partial^{*} v} X_e,\quad Z_p = \bigotimes_{e\in\partial p} Z_e,
\end{equation}
where $v$, $e$, and $p$ denote the vertices, edges, and faces of the square lattice, respectively. Here, $\partial$ and $\partial^{*}$ are the boundary operators of the square lattice and its dual lattice. The boundaries of the code are classified into two types: the smooth boundary, where $X_s$ appears at the edge, and the rough boundary, where $Z_p$ appears at the edge.

\subsection{Error Detection and Decoding}\label{subsec:measurement_decode}

The encoded state $\ket{\psi_L}$ is stabilized by the group $\mathcal{\stabilizergroup}$ as constructed above. Errors in this state are detected by measuring the eigenvalues corresponding to the stabilizer generators $S \in \stabilizergenerators$, yielding either $+1$ or $-1$, which provides information about the Pauli errors that have occurred in the qubit array.
This information is obtained from indirect measurements of the star (plaquette) operators as shown in \figref{fig:syndrome_circuit} using the orange (blue) ancilla qubits in \figref{fig:SurfaceCode}.
This process is called the syndrome measurement.

As shown in~\figref{fig:time-dependent-erasure}, for a surface code with distance $d$, we perform more than $d$ rounds of syndrome measurements. Even if measurement errors occur during these syndrome measurements, repeating them $d$ times ensures that the logical qubit is robust against measurement errors.

The error syndromes contain information about the errors. A matching graph $\mathcal{N}$, in which vertices represent spatiotemporal flips of error syndromes from $+1$ to $-1$ or vice versa~\cite{Fowler2012-hs, Delfosse2017-be, IOlius2023-ty, higgott2023sparse}, is formed to decode these syndromes into Pauli errors. The detector error model is a typical method to form $\mathcal{N}$ from syndrome measurement circuits~\cite{gidney2021stim}.
Then, a Minimum Weight Perfect Matching (MWPM) problem solver~\cite{Kolmogorov2009-eb, higgott2023sparse} matches the error-detected vertices to find the most likely error chains.

\subsection{Previous Work on Erasure Errors}

Erasure errors can be classified into three categories: static erasure errors, dynamic erasure errors that can be corrected immediately, and dynamic erasure errors that cannot be corrected immediately.


In systems where erasure errors occur statically, the positions and number of erasure errors are known before executing quantum computation, and they remain unchanged during execution. A typical case of static erasure errors is fabrication defects in superconducting quantum circuits~\cite{Nagayama2017-gq, auger2017fault}. 
%
This type of erasure can be corrected using \emph{super stabilizers}. There are two main methods for forming super stabilizers: the method of Stace and Barrett~\cite{Stace2010-ka} and the method of Nagayama \textit{et al.}~\cite{Nagayama2017-gq}. The first method is simple and easy to implement; super stabilizers are created by multiplying stabilizers associated with the erased data qubits, as illustrated in \figref{fig:SurfaceCode}. 
%

Some systems may experience erasure errors dynamically during computation, where the number of erasure errors changes over time. Mechanisms for detecting the positions of erased qubits have been investigated, depending on physical and architectural systems~\cite{Whiteside2014-rc, Cong2021-bj, Wu2022-uy, Chow2024-fb}. 
There are two types of such dynamic erasure errors. Erasure errors of the first type remain unless corrected.
A typical example of this type is neutral atom quantum computers; erasure errors remain unless special operations are executed to fix them, such as refilling the erased position with a reservoir atom instantly. This refilling is achieved by transporting atoms using optical tweezers. In the erasure conversion approach~\cite{Wu2022-uy, Sahay2023-av}, the original qubit information is exactly erased, but the atom may remain in its position.
In this case, the erased qubits, which are in a completely mixed state, can be incorporated into the surrounding stabilizers and projected onto the proper codeword state of the quantum error-correcting code. 



Some systems cannot correct dynamic erasure sequentially after detecting their occurrence~\cite{McEwen2021-zn}. 
In dense arrays of neutral atom quantum computers, sequential erasure correction will be difficult because transporting atoms into the online array can disturb other qubits~\cite{Barredo2016-sb}. Consequently, the number of erasure errors in the array increases over time. In such cases, the surface code does not function properly because the distribution of erased qubits changes continuously, and error detection and decoding do not work effectively around erased qubits.

\section{Pauli Error Correction under Accumulated Dynamic Erasure Errors}

\subsection{Main Problem}

We consider a situation where erasure errors occur dynamically and cannot be corrected immediately after detection. Neutral atom quantum computers with dense optical tweezer arrays correspond to this scenario since atoms cannot be transported to the dense online array on which logical qubits are encoded because atom transport possibly causes adverse effects such as decoherence or repulsion between atoms.

\subsection{Generating Stabilizer Generators}\label{subsec:superstabilizer}

When erasure errors occur on data qubits, we can maintain the error-correcting functionality by modifying the stabilizer group $\mathcal{S}=\langle\mathcal{G}\rangle$. This is achieved by removing the stabilizer generators associated with the erased data qubits and adding super star (or plaquette) operators to the stabilizer group by multiplying the removed stabilizer generators.

Here, we use Stace’s method to compose stabilizers into super stabilizers. Suppose a data qubit has been erased, as shown in the center of \figref{fig:SurfaceCode}. Two star stabilizer generators act on this erased data qubit; let’s call them $X_{i}, X_{j} \in \stabilizergenerators$. To exclude the erased data qubit from the code, we remove $X_i$ and $X_j$ from $\stabilizergenerators$ and add the merged stabilizer
\begin{equation}
X_{ij} = X_{i} X_{j}
\end{equation}
to $\stabilizergenerators$.

Similarly, on the plaquette stabilizer side, two stabilizers share the erased data qubit, and we can compose them in the same way. Even when multiple erasures exist, we can generate the deformed stabilizer group by inserting stabilizers created by multiplying the stabilizers that share the erased data qubits. We denote the new generators formed in this way as $\tilde{\stabilizergenerators}$. The stabilizer group $\tilde{\stabilizergroup} = \langle\tilde{\stabilizergenerators}\rangle$, which has $\tilde{\stabilizergenerators}$ as its generators, stabilizes the surface code state of the qubit array containing erasures.

\subsection{Composition of Super Stabilizers and Decoding under Dynamic Erasure}

Suppose dynamic erasure errors occur in an atom array during the execution of stabilizer circuits. In such cases, erasure detection should be performed before measuring syndromes for all stabilizer generators. This allows for constructing super stabilizers, which exclude erased qubits from the surface code. Based on the distribution of detected erased qubits, we construct the super stabilizers as described above and then perform syndrome measurements corresponding to the stabilizer generators that include these super stabilizers.

We update the stabilizer generators before each full round of syndrome measurements to adapt to dynamically occurring erasure errors. We must detect erasure errors and their distribution on the array in every syndrome measurement round to compose the new stabilizer generators with the procedure described in Section~\ref{subsec:superstabilizer}. In this study, we apply the Union-Find algorithm~\cite{Kozen1992} for finding the stabilizer pairs following the erasure detection, as described in Algorithm~\ref{algo:UF-stab}.

Then, the matching graph $\mathcal{N}$ is formed from the syndromes that are measured with the updated stabilizer generators.
Finally, an MWPM solver decodes the syndromes, and we achieve Pauli error correction in the presence of erasure errors.

\begin{algorithm}[H]
    \caption{The method to compose super stabilizers via Union Find algorithm}
    \label{algo:UF-stab}
    \begin{algorithmic}[1]
    \Function{union\_find}{$stabilizers$}
        \State $uf\gets UnionFindStructure(stabilizers)$
        \ForAll{$stabilizer0 \gets stabilizers$}
            \State $data\_qubits \gets stabilizer.data\_qubits$
            \ForAll{$qubit \gets data\_qubits$}
                \If{$qubit\;is\;erased$}
                \State $stabilizer1 \gets Find\_another\_stabilizer(qubit)$
                \State $uf.union(stabilizer0, stabilizer1)$
                \EndIf
            \EndFor
        \EndFor
        \State \Return $uf$
    \EndFunction
    
    \Function {create\_stabilizers}{$ideal\_stabilizers$}
        \State $uf\gets \Call{union\_find}{ideal\_stabilizers}$
        \State $roots\gets uf.roots()$
        \State $new\_stabilizers\gets \Call{merge\_stabilizers}{root, ideal\_stabilizers}$
        \Comment{This function merges stabilizers belonging to the same tree using Stace's method.}
        \State \Return $new\_stabilizers$
    \EndFunction
    \end{algorithmic}
\end{algorithm}

\section{Numerical Simulation and Analysis}

\subsection{Error Models and Assumptions in Our Simulation}

We performed circuit-based Monte Carlo simulations to investigate the behavior of logical failures and catastrophic corruption of the array. We used Stim~\cite{gidney2021stim} and Pymatching~\cite{higgott2023sparse} to conduct all numerical experiments. 

We assume a neutral atom device that can perform site-selective gate operations and non-destructive, mid-circuit measurements of qubits in situ.
We considered two types of errors occurring in physical qubits: Pauli errors, which occur within the qubit space, and erasure errors. 

We assumed that the Pauli errors are depolarizing errors. The depolarizing error channel on a single qubit is defined as an error channel where the three Pauli operators $X$, $Y$, and $Z$ are applied with equal probability $\pdep/3$, i.e.,
\begin{equation}
\mathcal{E}(\rho) = (1-\pdep)\rho +\frac{\pdep}{3}\sum_{M\in {X,Y,Z}}
M\rho M^{\dagger}.
\end{equation}
Similarly, the depolarizing error on two qubits with probability $\pdep$, typically caused along with two-qubit gates, is defined as
\begin{equation}
\mathcal{E}(\rho) = (1-\pdep)\rho +\frac{\pdep}{15}\sum_{M\in \mathcal{P}} M\rho M^{\dagger},
\end{equation}
where the set $\mathcal{P}$ is given by
\begin{equation}
\begin{split}
\mathcal{P} = \{ &IX, IY, IZ,
XI, XX, XY, XZ,\\
&YI, YX, YY, YZ,
ZI, ZX, ZY, ZZ
\}.
\end{split}
\end{equation}

Erasure errors occur with probability $\perasure$ before each round of syndrome measurements, and they remains on the array once an erasure occurs.
We suppose that erasure errors can be detected simultaneously after extracting error syndromes without fail or errors for simplicity, i.e. the erasure errors affect a single round of syndrome extraction, and we can know the erased qubits after the measurement.
We further assume that erasure errors are correctable only if the atom array does not store information of any logical qubit encoded by the surface code, i.e., only incoherent recovery of erased qubits is allowed.
This erasure error model causes the accumulation of erasure errors.
To simulate these accumulated erasure errors in Stim, we generated $10^4$ instances of random erasure occurrence over up to $10d$ rounds of syndrome measurements. Each instance represents the timings and locations of random erasure events obeying an erasure probability.
If the instance indicates the erasure occurrence, the completely depolarizing channel 
\begin{equation}
    \mathcal{E}(\rho) = \frac{I}{2},
\end{equation}
is applied to the erased qubits.
We model this process in the simulation by applying a uniformly random Pauli error from $\{I, X, Y, Z\}$~\cite{Fowler2013-sr,Wu2022-uy}.

\subsection{Composition of Super Stabilizers}\label{subsec:superstabilizer_composition}

The set of stabilizer generators is composed based on detected erased qubits until the previous measurement rounds, before each syndrome measurement round. In this study, we formed super stabilizers by Stace’s method. The sets of stabilizer generators are formed into super stabilizers by using a Union-Find data structure, as shown in the pseudocode in Algorithm~\ref{algo:UF-stab}. We can execute this algorithm to generate stabilizer generators for a system with erasures in $O(\log N)$ time steps for $N$ qubits.

The composition of stabilizer generators reduces the effective code distance. The length of the shortest logical operator (composed of a corresponding chain of physical operators) that traverses the larger, composed stabilizer generator becomes shorter than the shortest logical operator in the usual surface code with no erasures. The existence of a shorter logical operator results in a decrease in the effective code distance.

\begin{figure}[htbp]
\centering
\includegraphics[width=0.8\linewidth]{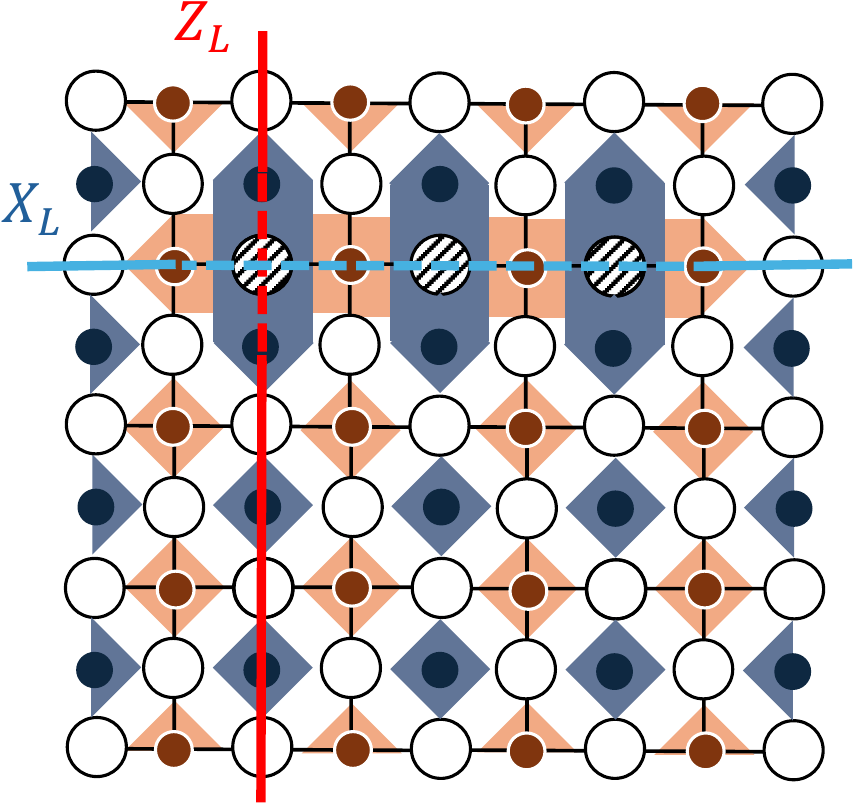}
\caption{A surface code of code distance $d=5$ with erasure errors, which include shortened logical operators due to super stabilizers. The blue and red lines represent the shortest $X$ and $Z$ logical operators, $X_L$ and $Z_L$. The $X_L$ is shortened to $d=3$, and the $Z_L$ is shortened to $d=2$ because super stabilizers adapt to erasure errors. The $Z_L$ of $d=2$ no longer has tolerance against $Z$ errors.}
\label{fig:code_distance}
\end{figure}

As shown in the following section and in \figref{fig:effective_code_distance}, performing multiple rounds of syndrome measurements induces a shorter effective code distance because erasures accumulate along with the rounds. Suppose erasure errors occur continuously between the boundaries, as in \figref{fig:code_distance}, a larger stabilizer generator that connects the boundaries will be composed. 
The effective code distance of this lattice is $2$, and it does not have sufficient code distance to protect a logical qubit against Pauli errors. Therefore, if erasure errors span from a boundary to the opposite boundary, we declare the logical qubit destroyed in this study.

\subsection{Measurement of Error Syndromes and Decoding Pauli Errors}

We generate a syndrome measurement circuit for each random erasure instance to be executed in our simulation.
For a surface code with distance $d$, we perform $d$ rounds of syndrome measurements with erasure detection performed before each syndrome measurement. 
Erasure detections are supposed to always succeed immediately without performing a dedicated erasure-detection sequence.

After erasure detections and syndrome measurements, we created the detector error model from the syndrome circuit and its matching graph $\mathcal{N}$, where vertices hold the flips of measurement results. We decoded the error syndromes in $\mathcal{N}$ into error locations using Pymatching.

\subsection{Numerical Results on a single array}\label{subsec:Numerical_Results}

\begin{figure*}
\centering
\includegraphics[width=0.9\linewidth]{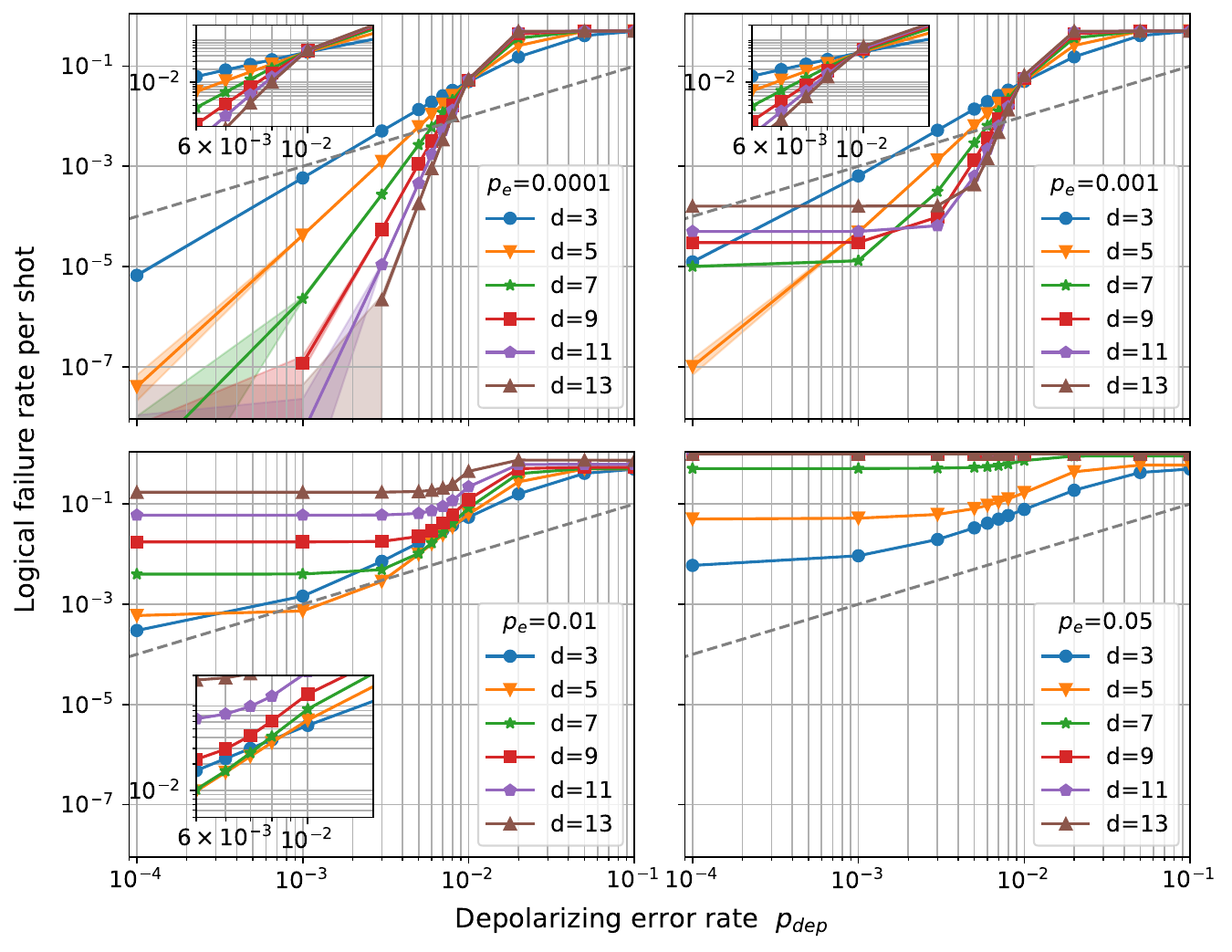}
\caption{The logical failure rate after $d$-rounds of syndrome measurements for each erasure error rate $p_{\textrm{e}}$. Each colored line represents a logical failure rate (vertical axis) vs. the depolarizing error rate per gate operation $p_{\textrm{dep}}$ (horizontal axis) for code distances $3$, $5$, $7$, $9$, $11$, and $13$. Each graph has a fixed erasure error rate $p_{\textrm{e}}$. The black dashed line is the break-even line, where the error correction suppresses errors below the depolarizing error rate.
Each inset is a zoom-in to the cross-point around $p_{\text{dep}}=10^{-2}$.}
\label{fig:erasure_threshold}
\end{figure*}

First, we calculated the threshold of the depolarizing error rate against the logical qubit error rate for each erasure error rate. ~\figref{fig:erasure_threshold} shows the results of sweeping $\pdep$ with fixed $\perasure$. 
Each data point in \figref{fig:erasure_threshold} is the average of the logical failure rate of the $l=10^4$ random erasure instances. Each instance simulation runs until $\xi_i=10^4$ logical failures are detected or until $\Omega_i\le 10^5$ attempts are completed, where $\Omega_i$ and $\xi_i$ are the number of sampling and the number of
logical failure events for each erasure instance, respectively. Hence, each data point is generated by $10^9$ attempts at maximum.
In general, logical failures are unintended flips of logical operators.
In this study, disruption of the error-correcting code due to erasure errors is also regarded as a failure.

Each shaded area in~\figref{fig:erasure_threshold} represents a confidence interval of the maximum-likelihood estimation of logical failure rate~$p^{*}={\xi}/{\Omega}$ under the assumption that the data follow a binomial distribution
\begin{equation}
f_{\Omega,\xi}(p)=\binom{\Omega}{\xi}\,p^{\xi}(1-p)^{\Omega-\xi},
\end{equation}
where $\Omega$ and $\xi$ are the total number of sampling and the number of logical failure events, respectively.
For this case the shaded interval $[p_{\text{low}},\,p_{\text{high}}]$
is defined as the region in which the likelihood falls to at most
$1/10^{3}$ of its maximum value, i.e.
\begin{equation}\label{eq:confidence_interval}
f_{\Omega,\xi}(p_{\text{low}})\;=\;
f_{\Omega,\xi}(p_{\text{high}})\;=\;
\frac{1}{10^{3}}\;f_{\Omega,\xi}(p^{*}) .
\end{equation}
In this simulation, we performed sampling until $\xi_i=10^4$ logical failures are detected or until $\Omega_i = 10^5$ attempts for each of $l = 10^{4}$ erasure instances. Hence the total number of sampling and the total number of logical failure events are 
\begin{equation}
    \qquad
    \xi \;=\; \sum_{i=1}^{l}\xi_i.
\end{equation}

\begin{figure*}
\centering
\includegraphics[width=0.9\linewidth]{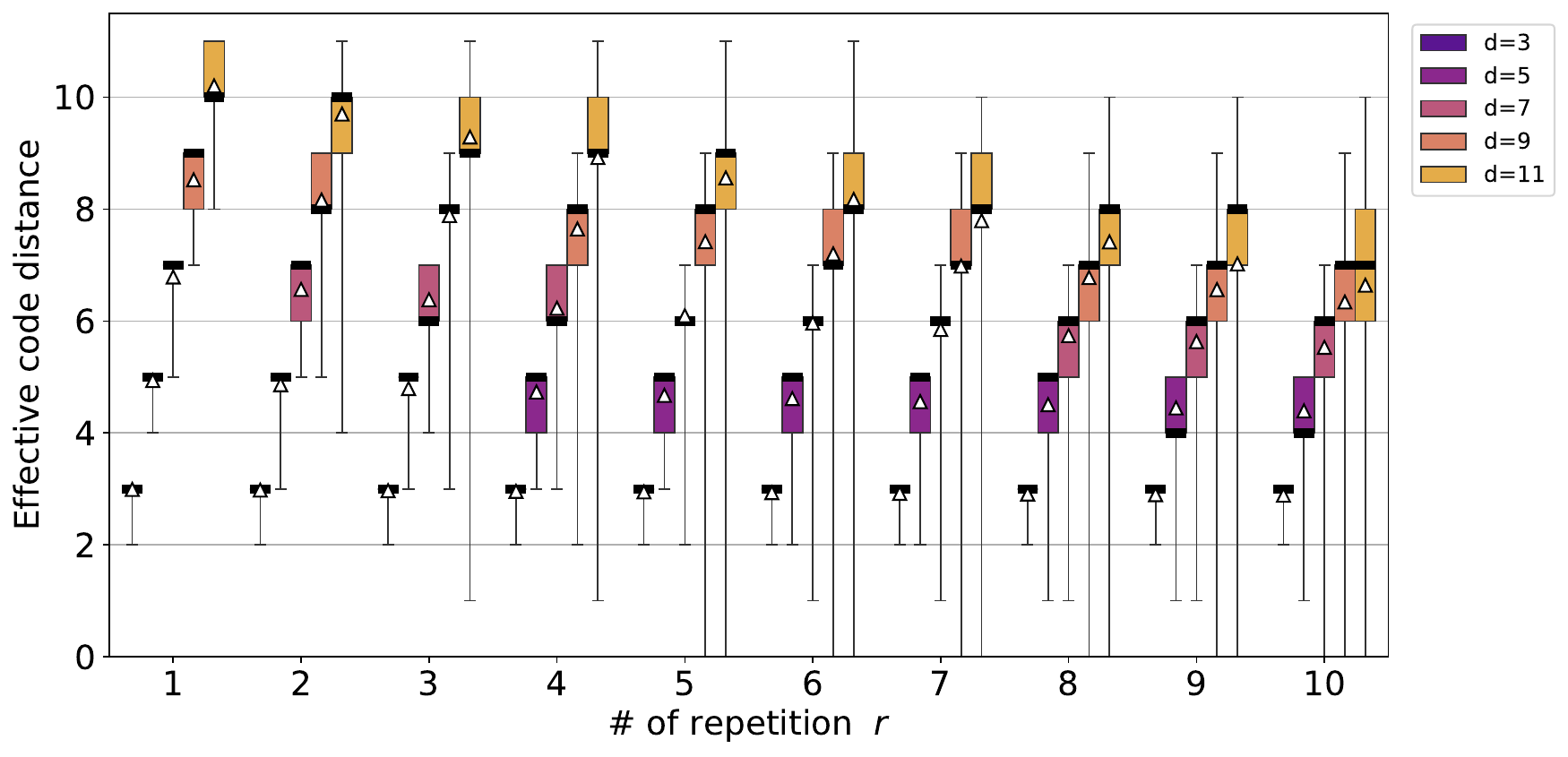}
\caption{This plot shows the decrease in effective code distances with increasing rounds of the syndrome measurements. The horizontal axis $r$ is the repetition number of syndrome measurements normalized by the code distance $d$, i.e., performing the syndrome measurement cycles $dr$ times. Each box plot is obtained from $10^4$ circuit samples with $p_{\textrm{e}} = 1.0 \times 10^{-3}$. The detector error model of every circuit provides the effective code distance as the “shortest graph-like error” in Stim~\cite{gidney2021stim}. Each thick black line is the median, and each white triangle is the average of the effective code distance. The whiskers of each box represent the maximum and minimum of the samples. 
Effective code distances of longer initial code distances decrease rapidly because they are more exposed to accumulated erasure errors due to $d$-repetitions.}
\label{fig:effective_code_distance}
\end{figure*}

~\figref{fig:erasure_threshold} shows how the erasure error rate affects the logical failure rate of the surface code.
Note that the codes with longer code distances require more repetitions of syndrome measurements, resulting in a higher erasure ratio at the end of the $d$ rounds.
As a result, codes with longer code distances degrade more significantly.
Due to the faster degradation of longer code distances, the plots for the higher erasure rates in ~\figref{fig:erasure_threshold} do not show the explicit threshold of the depolarizing error rate.
In each plot of erasure error rate $p_{\textrm{e}} = 1\times 10^{-4}$, $1\times 10^{-3}$ and $1\times 10^{-2}$, there is a threshold-like cross point around $p_{\textrm{dep}}=1\times10^{-2}$, but not all curves pass through this cross point; it is a pseudo-threshold.
It suggests that surface codes can reduce the logical failure rate by increasing the code distance up to a certain length, but further increasing the code distance beyond that length does not effectively suppress logical failures.
For instance, the graph for the erasure rate of $p_e = 1\times 10^{-3}$ shows that the surface code can suppress the logical failure rate when the depolarizing error rate is approximately between $3\times 10^{-3}\le p_{\textrm{dep}}\le 1\times10^{-2}$ by increasing the code distance up to $11$.
However, the logical failure rate for $d=13$ is worse than for the shorter code distances.
Other plots also suggest that long code distances increase the logical failure rate.
Thus, low erasure error rates are essential to successfully performing error correction with a longer code distance.

An analytical derivation also supports this result. The derivation indicates that the accumulated-erasure error model does not have a threshold, meaning $p_{\textrm{e, th}} = 0$; the ratio of qubits erased after repeating $d$ times of syndrome measurements is
\begin{equation}
R = 1 - (1 - p_{\textrm{e}})^{\alpha d},
\end{equation}
where $\alpha$ is the number of gates applied to a qubit.
This equation tells that $R \rightarrow 1$ in the limit of $d \rightarrow \infty$. 
This means that erasure errors cannot be attenuated, regardless of how much the code distance increases, as long as erasure errors accumulate.

Secondly, we also evaluated the evolution of the effective code distance of each syndrome measurement round, as shown in \figref{fig:effective_code_distance}. Here, we used the length of the shortest graph-like error, that is the shortest undetected logical error on the detector error model, in Stim~\cite{gidney2021stim} as the effective code distance. 
The effective code distance is granted zero if the erasure errors destroy the logical qubit.
This reduction of each code distance is caused by composing super stabilizers to adapt to erasure errors, as described in \secref{subsec:superstabilizer_composition}.
Notably, codes with longer initial code distances lose their effective code distance more rapidly. This is also because a longer code distance results in a larger erasure ratio at the end of the $r \times d$ rounds.

\begin{figure}[t]
\centering
\subfloat[Transferring a logical qubit]{
    \includegraphics[width=0.95\linewidth]{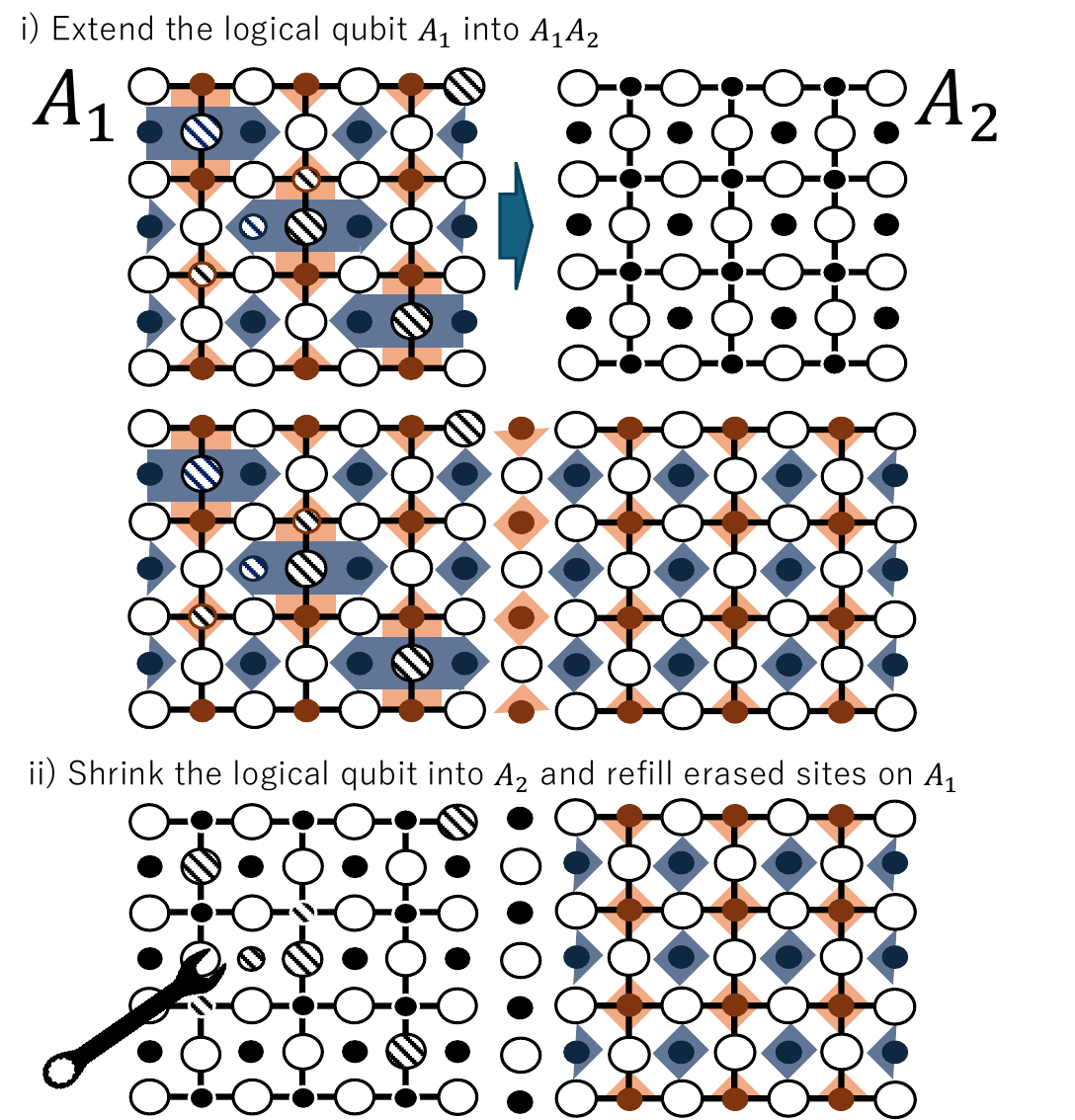}
}\\
\subfloat[Timeline of the 3-shift erasure recovery]{
    \includegraphics[width=0.95\linewidth]{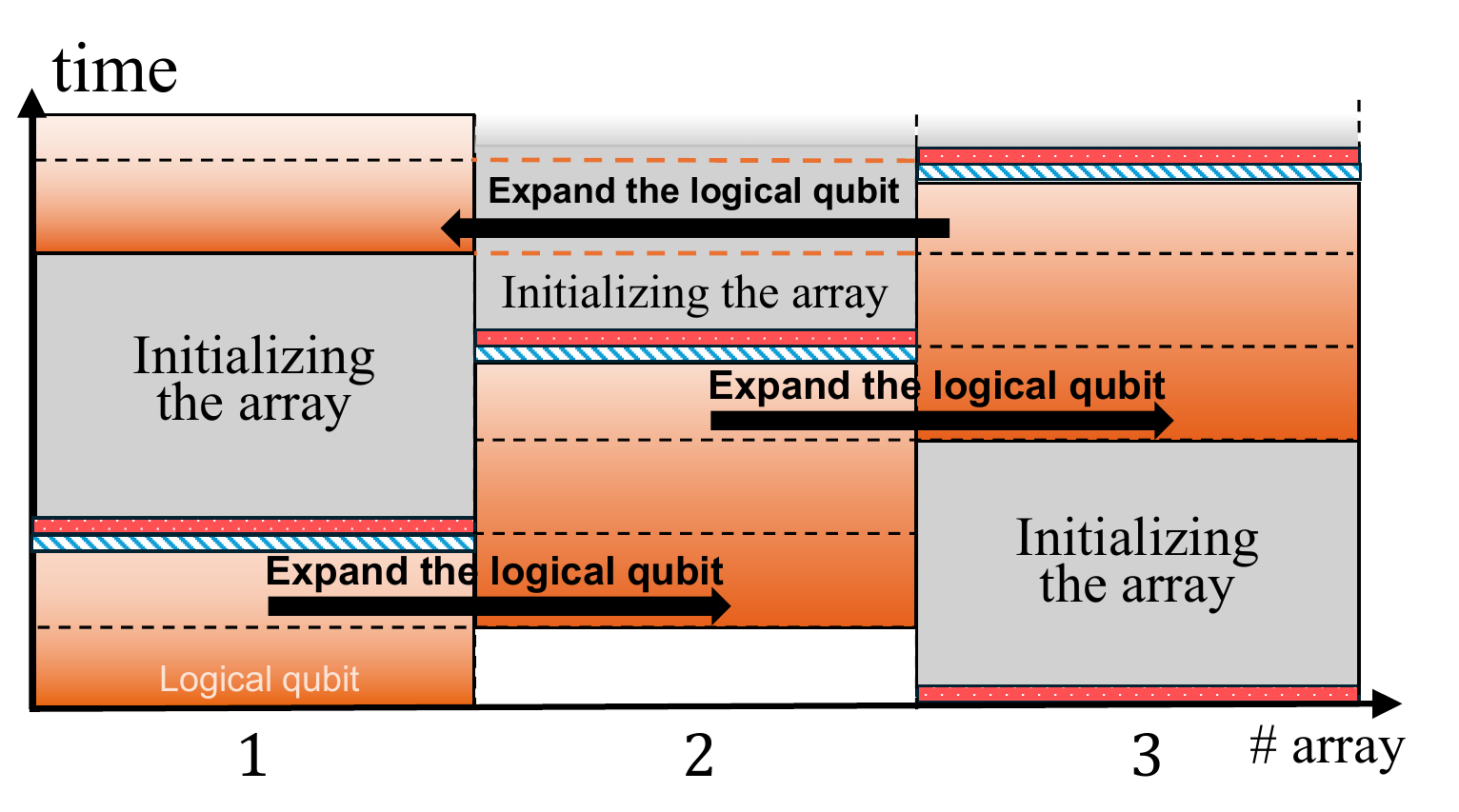}
}
\caption{
(a) The process of transferring a logical qubit via code deformation.
(b) The timeline of the 3-shift erasure recovery process. The horizontal axis represents the indexing of three arrays used to store the information of the logical qubit. The orange blocks indicate the location of the logical qubit, while the shaded blue blocks denote the measurement of data qubits performed to contract the extended logical qubit. The red blocks indicate the relocation of an array to the position where it will be utilized in the next step, depending on the geometric configuration of the tweezer arrays. 
Initially, the logical qubit is allocated to $\mathcal{A}_1$. Subsequently, the logical qubit is expanded to span $\mathcal{A}_1$ and $\mathcal{A}_2$. Following this, the logical qubit is contracted to $\mathcal{A}_2$ by measuring the data qubits on $\mathcal{A}_1$. Finally, $\mathcal{A}_1$ is shifted to the adjacent to $\mathcal{A}_3$, as $\mathcal{A}_1$ will receive the logical qubit from $\mathcal{A}_3$ in the subsequent step.
}
\label{fig:code_deformation}
\end{figure}

\section{$k$-Shift Erasure Recovery}

Our simulations demonstrated that repeating the stabilizer circuit causes more erasure accumulation and continuously decreases effective code distance. 
Eventually, the code loses its tolerance against depolarizing errors.

To address this problem, we propose an approach that moves the logical qubit from the atom array with accumulated erasure errors to a new atom array free from erased qubits.
We call this scheme \textit{$k$-shift erasure recovery}.
This approach can be applied to other types of quantum computers with similar problems, such as qubits getting lost or fidelities decreasing as computation processes in a manner difficult to fix while computation is ongoing.
Our approach is achieved by two steps, combining the expansion and contraction operations of the surface code through code deformation~\cite{Horsman2011-yv} (\figref{fig:code_deformation}(a)):

\begin{enumerate}
\item Repeat the surface code procedure, tolerating state errors, leakage errors, and erasure errors while executing quantum computation.
\item When the accumulation of erasure errors on the array threatens to reduce the error tolerance of the surface code below the required level, transfer the logical qubit to another perfect atom array. Then, refresh the old atom array that contains erased qubits to restore it to a defect-free array.
\end{enumerate}

This series of steps separates the functionality of executing quantum computation—protecting against state errors and enduring erasure errors—from the functionality of correcting erasure errors. By moving the logical qubit away from the imperfect array, our scheme prevents the logical qubit from suffering effects caused by repairing the atom array, such as noise or increased errors during the induced idling time.

Consider a qubit array $\mathcal{A}_1$ on which a logical qubit of the surface code is encoded and another qubit array $\mathcal{A}_2$ on which nothing is encoded initially. Then, syndrome measurements and error corrections on $\mathcal{A}_1$ are repeatedly performed to correct Pauli errors. However, this iteration increases the erasure ratio, eventually making the logical failure rate exceed acceptable levels (e.g., $10^{-15}$~\cite{Fowler2012-hz}).

Our protocol performs code deformation~\cite{Horsman2011-yv} to move the logical qubit from $\mathcal{A}_1$ to the new perfect array $\mathcal{A}_2$ before the logical failure rate becomes too high due to the accumulation of erased qubits. First, initialize the qubits in $\mathcal{A}_2$ appropriately to $\ket{0}$ or $\ket{+}$, depending on the direction of the boundary expansion. Next, perform $d$ rounds of syndrome measurements on $\mathcal{A}_1$ and $\mathcal{A}_2$ to merge them.
Then, the logical qubit is expanded from size $d \times d$ on $\mathcal{A}_1$ to size $d \times 2d$ spanning both $\mathcal{A}_1$ and $\mathcal{A}_2$.

After that, measure the data qubits on $\mathcal{A}_1$ in the $Z$ or $X$ basis, depending on the boundary direction again, to contract the logical qubit to size $d \times d$ on $\mathcal{A}_2$. From the decoding result of the syndromes and the outcomes of the direct measurements on the data qubits, the parity to be applied to the error correction of the logical operator is determined and recorded.

Now, the qubits in $\mathcal{A}_1$ no longer hold logical information. Therefore, we can perform any operations on these qubits, even destructive ones, without concerns. Thus, the imperfect array with erased qubits in $\mathcal{A}_1$ can be repaired by rearranging or refilling the array with single atoms from an atom reservoir. For neutral atom array quantum computers, optical tweezers have been demonstrated to work well for such precise movement and insertion of atoms~\cite{Barredo2016-sb}.

Our scheme separates the process of maintaining the coherence of logical qubits (through syndrome measurements) from repairing the array (through operations that might be destructive to quantum data, such as rearrangement and refilling atoms via tweezers). This allows for more aggressive rearrangement strategies, including incoherent transportation and completely reloading atoms, compared to methods that sequentially replenish erased qubits.

In the same way, erased qubits on the new array $\mathcal{A}_2$ will accumulate after repeating error correction. Again, we can make the system tolerant to erasure errors by transferring the logical qubit to another array $\mathcal{A}_3$ before the logical failure rate of $\mathcal{A}_2$ becomes too high due to accumulated erasure errors.
Repeating transferring the logical qubit $(k-1)$ times secures a sufficient time of the restoration of the array $\mathcal{A}_1$.
Thus, the logical qubit on $\mathcal{A}_{k}$ can be transferred onto $\mathcal{A}_{1}$.
Repetition of this cycle prevents logical qubit from a logical Pauli error and erasure errors.

The number of the redundant arrays $k$ is decided depending on the time needed to fix erasures on the array and the time until the array can no longer tolerate the accumulation of erasure errors and Pauli errors.
In this protocol, the number of arrays $k$ contributes to redundancy in the loss of information at the code-patch level.
Note that this redundancy is principally enough to use $k=2$ arrays if resorting of erased atoms is quickly accomplished.
However, if enough $k$ arrays are not prepared, this redundancy cannot be utilized. 
It is analogous to the bus error in classical computer architecture.
This events will cause the information loss of a logical qubit memory.
Therefore, in order to secure sufficient time to restore the atom arrays, this scheme introduces redundancy by increasing the number $k$ of arrays.

\begin{figure}[htb]
\centering
\includegraphics[width=0.95\linewidth]{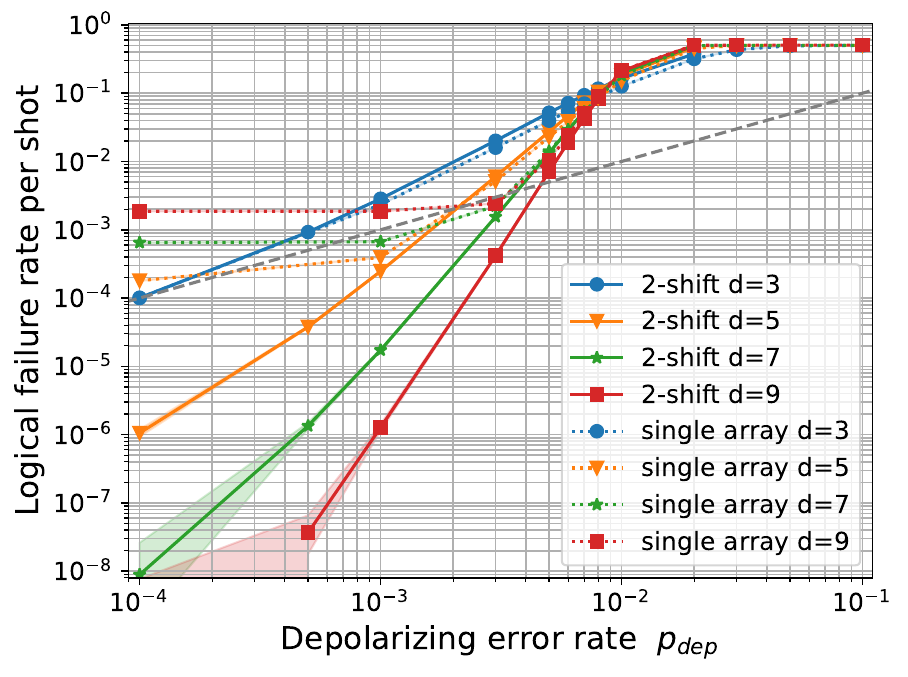}
\caption{Comparison of logical failure rates between conventional syndrome measurements and the $2$-shift erasure recovery for $p_{\textrm{e}} = 1.0 \times 10^{-3}$. The dotted lines represent the logical failure rate after $3d$ rounds of syndrome measurements on a single array, and the solid lines represent the $2$-shift erasure recovery that takes $3d$ rounds of syndrome measurements, where $d$ is the code distance. The black dashed line is the break-even line, where the error correction suppresses errors below the depolarizing error rate. For each code distance, the $2$-shift recovery performs the same or better logical failure rate, as long as the depolarizing error rate is lower than the threshold around $p_{\textrm{phys}} < 10^{-2}$. Each data point is obtained from $10^9$ attempts at maximum. Each shaded area represents the confidence interval of the sampling defined in~\eqref{eq:confidence_interval}. Note that the cross point of solid lines is a pseudo-threshold as the same as the single-patch simulation in~\figref{fig:erasure_threshold}. }
\label{fig:kshift_performance}
\end{figure}

To evaluate the performance of the $k$-shift recovery, the most important point is that transferring logical qubit with accumulated erasure errors effectively works. We numerically evaluated the effectiveness of the transferring logical qubits with accumulated erasure errors, i.e. $2$-shift erasure recovery. \figref{fig:kshift_performance} compares the logical failure rate between performing $3d$ rounds of syndrome measurements on a single array and performing the $2$-shift erasure recovery, which is the minimum setting of the $k$-shift erasure recovery. In the $2$-shift erasure recovery, we perform $d$ rounds of syndrome measurements on $\mathcal{A}_1$, then $d$ rounds on the expanded surface code spanning $\mathcal{A}_1$ and $\mathcal{A}_2$, followed by $d$ rounds of syndrome measurements on $\mathcal{A}_2$ after the contraction operation and measuring data qubits on $\mathcal{A}_1$ with Pauli Z-basis. Therefore, the "2-shift" data points are generated by simulations of 3 rounds in total. And "single array" data points are generated with 3 rounds of single arrays for comparison.
Pauli errors on the circuit are decoded by Pymatching with the detector error model of the above circuit.
Here, the vertical axis of \figref{fig:kshift_performance} represents the logical failure rate of measuring the final state on logical $Z$ basis after $3d$ rounds. Each data point is obtained from the average of the logical failure rate of the $10^4$ random erasure instances for $3d$ round measurements. As the same as Sec.~\ref{subsec:Numerical_Results}, each instance simulation runs until $10^4$ logical failures are detected or until $10^5$ attempts are completed. Hence, each data point is generated by $10^9$ attempts at maximum. Each shaded area represents the confidence interval of the maximum-likelihood estimation of logical failure rate in the same way shown in Sec.~\ref{subsec:Numerical_Results}.
These numerical results show that the $2$-shift recovery improves the logical failure rate compared to the simple repetition of syndrome measurements on a single array. For each code distance (except $d=3$), the $2$-shift recovery achieves suppression of the logical failure rate, whereas performing $3d$ rounds of syndrome measurements on a single array without shifting cannot suppress the logical failure rate at lower depolarizing error rates.
Shifting the logical qubit to the new array effectively resets the number of erased qubits that a logical qubit holds to zero. Thus, the continuous accumulation of erased qubits is prevented.
Therefore, $2$-shift erasure recovery can more effectively suppress the logical failure rate than single array cases.
This result suggests that $k$-shift recovery provides erasure-tolerance even in its smallest configuration since the proposed scheme works between patch $N$ and patch $N+1$ as well. 
Even if we conduct end-to-end simulations of such as $k=10$, failure rate per stabilizer round is considered the same, since erasure errors are refreshed at the beginning of each round.

We also calculate $k_{\mathrm{req}}$, the number of arrays required for a logical qubit to pass before returning to its original array so that a defective array has sufficient time to be restored to a perfect one.
The time scale to measure a physical qubit by imaging the fluorescence photons from a single atom is milliseconds~\cite{Norcia2023-zh}, while the time scale for gate operations is typically microseconds~\cite{PhysRevX.15.011009}.
Thus, the measurement on physical qubits is the most time-consuming operation in a syndrome measurement.
We assume that measuring the syndromes of all stabilizer generators in a single round takes $10$~ms, including erasure detection.
For the code with distance $d$, repeating $r$~rounds of syndrome measurements will consume $T_{\textrm{meas}}=10 rd$~ms.
On the other hand, sorting to a two-dimensional defect-free array also takes hundreds of milliseconds.
Suppose we have a square lattice array of $N=(2d-1)^2+2d$ atoms if the array is perfect.
In Ref.~\cite{Barredo2016-sb}, the time scale of sorting is estimated to be $0.28 N^{1.4}$~ms.
Therefore, the time scale to sort to a perfect array is $T_\textrm{sort}\propto O(d^{2.8})$.

\begin{figure}[htb]
    \centering
    \includegraphics[width=0.95\linewidth]{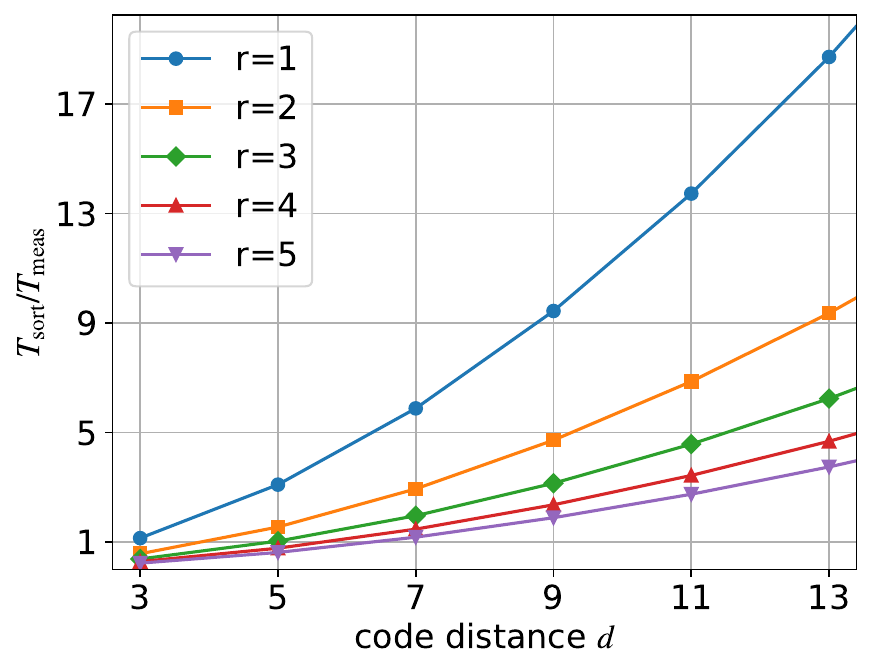}
    \caption{The dependency of $T_{\textrm{sort}}/T_{\textrm{meas}}$ for code distance on each number of repetition rounds~$r$.This ratio suggests the lower bound of $k$. This figure shows that the code with longer code distance requires more atom arrays.}
    \label{fig:estimation_k}
\end{figure}

$k_{\mathrm{req}}$ is lower bounded by the ratio of $T_{\textrm{meas}}$ and $T_{\textrm{sort}}$ since larger $k$ in this scheme allow longer time to recover erasures on the array.
\figref{fig:estimation_k} shows the ratio of $T_{\textrm{sort}}/T_{\textrm{meas}}$. 
Here, $r$ is the number of repeated rounds of error correction cycles where each cycle consists of $d$ syndrome measurements performed on the logical qubit in a given array before transferring to the next array.
For example of $r=2$, $d$ rounds of syndrome measurements are performed on a single array, then $d$ rounds of syndrome measurements are performed on a $d\times 2d$ array before contracting data qubits on the array containing erased qubits.
The figure suggests that a longer code distance requires that a logical qubit is transferred over more arrays.
When $d=9,\ r=2$ which is the largest setting in the numerical simulation in \figref{fig:kshift_performance}, $T_{\textrm{sort}}/T_{\textrm{meas}}\simeq 4.72$. 
Therefore, $k=5$ arrays are needed to protect a logical qubit.
Conversely, $k_{\mathrm{req}}$ decreases if the erasure rate allows more error correction cycles at each array.
Suppose an experimentalist sets the threshold of erased qubit ratio 
\begin{equation}
    R_{\mathrm{th}} = 1-(1-p_{\mathrm{e}})^{\alpha rd}.
\end{equation}
$k_{\mathrm{req}}$ is lower bounded by 
\begin{equation}
    k_{\mathrm{req}} \ge T_{\textrm{sort}}/T_{\textrm{meas}} \propto \left[(2d-1)^2+2d\right]^{1.4}\frac{\log(1-p_{\mathrm{e}})}{\log(1-R_\mathrm{th})}.
\end{equation}
Thus, $k_{\mathrm{req}}$ is minimized with respect to the code distance and the erasure error rate.

\section{Discussion and Conclusion}

In this paper, we have numerically verified the error correction performance of surface codes on systems where erasure errors continuously occur and accumulate on the array, such as in neutral atom quantum computers using optical tweezer arrays. 
Our work revealed that the surface code with longer code distances is more strongly affected by the accumulated erasure errors under the model where refilling erased qubits is impossible.
The logical failure rate becomes extremely high by increasing the code distance and repeating error correction because the threshold for the erasure error rate is theoretically zero.
Additionally, iterating the syndrome measurements increases the number of erased qubits. These erased qubits shorten the effective code distance due to the formation of super stabilizers, which worsens the fault tolerance.

To address the problem of error correction becoming impossible due to the accumulation of erased qubits, we proposed the \textit{$k$-shift erasure recovery} scheme to protect the logical qubit from erasure errors by transferring it to a perfect array using code deformation. We numerically demonstrated that the $k$-shift erasure recovery improves erasure tolerance compared to the simple repetition of syndrome measurements on a single array.

Our scheme allows us to rearrange physical qubits on offline arrays without concern about the side effects of rearranging atoms on the array due to separating the rearrangement of qubits from the coherence-preserving operations on logical qubits, including the Pauli error correction and logical qubit operations.
Our scheme enables more aggressive rearrangement strategies than instantly refilling reservoir atoms into the online array. For example, we can completely initialize the faulty offline array—discard all atoms and reload them again—instead of refilling defective sites by transporting atoms.
This approach works as an alternative to the coherent transportation of atoms, ensuring erasure tolerance, especially in systems where adaptive transportation to correct erasures is challenging or where the array needs refreshing due to erasures occurring more frequently than the capacity of atom transport.

In the following, we discuss aspects not considered in this study and open problems. The numerical calculations were performed under the assumption that the erasure error happens with the uniform rate before each syndrome measurement round. However, neutral atom quantum computers are prone to leakage errors and atom loss, mainly when performing two-qubit gates~\cite{Wu2022-uy, Sahay2023-av}.
Moreover, erased qubits should be detected with faulty components such that some leakage may spread before erasure conversion, while we assumed a noiseless erasure detection.
Therefore, investigating the effect of biased erasure error occurrence at mid-syndrome measurement rounds and faulty erasure detections are a subject for future work.

During syndrome extraction, ancilla qubits are also exposed to erasure errors in the simulation.
While our model did not address mitigating erasure errors in ancilla qubits, their effect is possibly mitigated by reusing other ancilla qubits of surrounding stabilizers. By reinitializing ancilla qubits before each use, they can be shared among multiple stabilizers.
Therefore, such a sharing strategy will further suppress the logical failure rate.

Investigating efficient scheduling and allocation of ancilla qubits for syndrome collection under this scheme should also be interesting. A compilation proposal that gradually changes the qubit allocations on the atom array has been proposed by~\cite{Baker2021-oc}. By adopting such compilation of scheduling and routing approaches with syndrome measurement circuits, the Rydberg atoms should realize syndrome collection with more efficient circulation of ancilla qubits and more efficient scheduling of syndrome measurements than recent studies that assume only nearest-neighbor interactions~\cite{Nagayama2017-gq}.
Researching the scheduling with the ancilla-sharing strategy mentioned above will promote the feasible implementation of the super stabilizers under accumulated erasure errors.
Moreover, a numerical simulation for repeating the transferring a logical qubit many times will reveal more detailed effect of $k$-shift erasure recovery for the logical error rate. However, we could not address this simulation because of the limitations of computational resources and runtime issues to perform such simulations. Providing the simulation of our framework on an efficient QEC simulation with erasure errors such as tensor-netowrk simulation\cite{Manabe2023-va} and Pauli+~\cite{Google-Quantum-AI2023-yd} is a future work.

We decoded error syndromes using Pymatching, an MWPM decoder. Recently, various improved decoders, such as the Union-Find decoder~\cite{Delfosse2017-be}, have been proposed. The Union-Find decoder is an almost linear time decoder for correcting Pauli and erasure errors. Research into utilizing such decoders instead of MWPM is valuable for achieving erasure tolerance in neutral atom quantum computers~\cite{Wu2022-uy}. It would be interesting to compare their performance for dynamically accumulating erasure errors.

We could not analytically derive the explicit effect of the accumulated erasure errors on decoding, which remains an open problem.
The accumulated erasure errors can be modeled as temporally correlated, completely depolarizing errors.
Evaluating these errors by mapping to the statistical mechanics model with long-range correlations~\cite{Chubb2021-in} is expected to reveal the effect of accumulated erasure errors on the logical failure rate. 

We exemplified our proposal using code deformation to transfer logical qubits from an imperfect array due to erasure errors to a new, perfect array. It would be interesting to compare our scheme of code deformation with other methods, such as those using quantum teleportation or transversal swap gates, since Clifford gates can be performed transversely on surface codes. These methods may transfer logical qubits more efficiently than the code deformation employed in our numerical simulations.

We would like to generalize our method to repair more faults on quantum computers and to ensure the sustainability of quantum information processing. Our scheme can be applied to other faults beyond erasure correction, leveraging quantum networks and distributed quantum computing—much like classical cloud computing moves or relocates services and data from one computer to another to sustain services permanently. In this context, $k$-shift recovery can recover faults on other types of quantum devices, such as superconducting circuits and trapped ion systems, as long as these devices are modularized. Our scheme enables the preservation of logical qubits from critical faults by transferring them from a faulty quantum device to a fresh one. Such distributed maintainability will be fundamental for the reliable and continuous use of fault-tolerant quantum computers.

\section*{Acknowledgment}
The authors thank Thomas M. Stace, Takashi Yamamoto, Rikizo Ikuta, and Toshiki Kobayashi for their helpful advice.
The authors also thank Kentaro Teramoto for his kind help in conducting the data.
This research was supported by JST Moonshot R\&D (JPMJMS2066, JPMJMS226C).
FK was supported by JST, the establishment of university fellowships towards the creation of science technology innovation, Grant Number JPMJFS2125.
FK and SN acknowledge the members of Quantum Internet Task Force for comprehensive discussions of quantum networks.

\bibliographystyle{IEEEtran}
\bibliography{reference}

\providecommand{\noopsort}[1]{}\providecommand{\singleletter}[1]{#1}%
\begin{thebibliography}{10}
\providecommand{\url}[1]{#1}
\csname url@samestyle\endcsname
\providecommand{\newblock}{\relax}
\providecommand{\bibinfo}[2]{#2}
\providecommand{\BIBentrySTDinterwordspacing}{\spaceskip=0pt\relax}
\providecommand{\BIBentryALTinterwordstretchfactor}{4}
\providecommand{\BIBentryALTinterwordspacing}{\spaceskip=\fontdimen2\font plus
\BIBentryALTinterwordstretchfactor\fontdimen3\font minus
  \fontdimen4\font\relax}
\providecommand{\BIBforeignlanguage}[2]{{%
\expandafter\ifx\csname l@#1\endcsname\relax
\typeout{** WARNING: IEEEtran.bst: No hyphenation pattern has been}%
\typeout{** loaded for the language `#1'. Using the pattern for}%
\typeout{** the default language instead.}%
\else
\language=\csname l@#1\endcsname
\fi
#2}}
\providecommand{\BIBdecl}{\relax}
\BIBdecl

\bibitem{Bluvstein2023-sm}
D.~Bluvstein, S.~J. Evered, A.~A. Geim, S.~H. Li, H.~Zhou, T.~Manovitz,
  S.~Ebadi, M.~Cain, M.~Kalinowski, D.~Hangleiter, J.~P.~B. Ataides,
  N.~Maskara, I.~Cong, X.~Gao, P.~S. Rodriguez, T.~Karolyshyn, G.~Semeghini,
  M.~J. Gullans, M.~Greiner, V.~Vuleti{\'c}, and M.~D. Lukin,
  ``\BIBforeignlanguage{en}{Logical quantum processor based on reconfigurable
  atom arrays},'' \emph{\BIBforeignlanguage{en}{Nature}}, Dec. 2023.

\bibitem{Young2020-st}
A.~W. Young, W.~J. Eckner, W.~R. Milner, D.~Kedar, M.~A. Norcia, E.~Oelker,
  N.~Schine, J.~Ye, and A.~M. Kaufman,
  ``\BIBforeignlanguage{en}{Half-minute-scale atomic coherence and high
  relative stability in a tweezer clock},''
  \emph{\BIBforeignlanguage{en}{Nature}}, vol. 588, no. 7838, pp. 408--413,
  Dec. 2020.

\bibitem{Jenkins2022-im}
A.~Jenkins, J.~W. Lis, A.~Senoo, W.~F. McGrew, and A.~M. Kaufman,
  ``\BIBforeignlanguage{en}{Ytterbium nuclear-spin qubits in an optical tweezer
  array},'' \emph{\BIBforeignlanguage{en}{Phys. Rev. X.}}, vol.~12, no.~2, May
  2022.

\bibitem{Klusener2024-yu}
V.~Klüsener, S.~Pucher, D.~Yankelev, J.~Trautmann, F.~Spriestersbach,
  D.~Filin, S.~G. Porsev, M.~S. Safronova, I.~Bloch, and S.~Blatt,
  ``\BIBforeignlanguage{en}{Long-lived coherence on a $\mu$hz scale optical
  magnetic quadrupole transition},'' \emph{\BIBforeignlanguage{en}{Phys. Rev.
  Lett.}}, vol. 132, no.~25, p. 253201, Jun. 2024.

\bibitem{Ma2022-dt}
S.~Ma, A.~P. Burgers, G.~Liu, J.~Wilson, B.~Zhang, and J.~D. Thompson,
  ``\BIBforeignlanguage{en}{Universal gate operations on nuclear spin qubits in
  an optical tweezer array of {Yb171} atoms},''
  \emph{\BIBforeignlanguage{en}{Phys. Rev. X.}}, vol.~12, no.~2, May 2022.

\bibitem{Barredo2016-sb}
D.~Barredo, S.~de~L{\'e}s{\'e}leuc, V.~Lienhard, T.~Lahaye, and A.~Browaeys,
  ``\BIBforeignlanguage{en}{An atom-by-atom assembler of defect-free arbitrary
  two-dimensional atomic arrays},'' \emph{\BIBforeignlanguage{en}{Science}},
  vol. 354, no. 6315, pp. 1021--1023, Nov. 2016.

\bibitem{Barredo2018-ts}
D.~Barredo, V.~Lienhard, S.~de~L{\'e}s{\'e}leuc, T.~Lahaye, and A.~Browaeys,
  ``\BIBforeignlanguage{en}{Synthetic three-dimensional atomic structures
  assembled atom by atom},'' \emph{\BIBforeignlanguage{en}{Nature}}, vol. 561,
  no. 7721, pp. 79--82, Sep. 2018.

\bibitem{Bluvstein2022-ll}
D.~Bluvstein, H.~Levine, G.~Semeghini, T.~T. Wang, S.~Ebadi, M.~Kalinowski,
  A.~Keesling, N.~Maskara, H.~Pichler, M.~Greiner, V.~Vuleti{\'c}, and M.~D.
  Lukin, ``\BIBforeignlanguage{en}{A quantum processor based on coherent
  transport of entangled atom arrays},''
  \emph{\BIBforeignlanguage{en}{Nature}}, vol. 604, no. 7906, pp. 451--456,
  Apr. 2022.

\bibitem{gottesman1997stabilizer}
D.~Gottesman, \emph{Stabilizer codes and quantum error correction}.\hskip 1em
  plus 0.5em minus 0.4em\relax California Institute of Technology, 1997.

\bibitem{Preskill1997-ii}
J.~Preskill, ``Fault-tolerant quantum computation,'' \emph{arXiv [quant-ph]},
  Dec. 1997.

\bibitem{Saffman2016-br}
M.~Saffman, ``\BIBforeignlanguage{en}{Quantum computing with atomic qubits and
  rydberg interactions: progress and challenges},''
  \emph{\BIBforeignlanguage{en}{J. Phys. B At. Mol. Opt. Phys.}}, vol.~49,
  no.~20, p. 202001, Oct. 2016.

\bibitem{Chow2024-fb}
\BIBentryALTinterwordspacing
M.~N.~H. Chow, V.~Buchemmavari, S.~Omanakuttan, B.~J. Little, S.~Pandey, I.~H.
  Deutsch, and Y.-Y. Jau, ``Circuit-based leakage-to-erasure conversion in a
  neutral-atom quantum processor,'' \emph{PRX Quantum}, vol.~5, p. 040343, Dec
  2024. [Online]. Available:
  \url{https://link.aps.org/doi/10.1103/PRXQuantum.5.040343}
\BIBentrySTDinterwordspacing

\bibitem{Wu2022-uy}
Y.~Wu, S.~Kolkowitz, S.~Puri, and J.~D. Thompson,
  ``\BIBforeignlanguage{en}{Erasure conversion for fault-tolerant quantum
  computing in alkaline earth rydberg atom arrays},''
  \emph{\BIBforeignlanguage{en}{Nat. Commun.}}, vol.~13, no.~1, p. 4657, Aug.
  2022.

\bibitem{Ma2023-jp}
S.~Ma, G.~Liu, P.~Peng, B.~Zhang, S.~Jandura, J.~Claes, A.~P. Burgers,
  G.~Pupillo, S.~Puri, and J.~D. Thompson,
  ``\BIBforeignlanguage{en}{High-fidelity gates and mid-circuit erasure
  conversion in an atomic qubit},'' \emph{\BIBforeignlanguage{en}{Nature}},
  vol. 622, no. 7982, pp. 279--284, Oct. 2023.

\bibitem{Scholl2023-gw}
P.~Scholl, A.~L. Shaw, R.~B.-S. Tsai, R.~Finkelstein, J.~Choi, and M.~Endres,
  ``\BIBforeignlanguage{en}{Erasure conversion in a high-fidelity rydberg
  quantum simulator},'' \emph{\BIBforeignlanguage{en}{Nature}}, vol. 622, no.
  7982, pp. 273--278, Oct. 2023.

\bibitem{Stace2010-ka}
T.~M. Stace and S.~D. Barrett, ``Error correction and degeneracy in surface
  codes suffering loss,'' \emph{Phys. Rev. A}, vol.~81, no.~2, p. 022317, Feb.
  2010.

\bibitem{Nagayama2017-gq}
S.~Nagayama, A.~G. Fowler, D.~Horsman, S.~J. Devitt, and R.~Van~Meter,
  ``\BIBforeignlanguage{en}{Surface code error correction on a defective
  lattice},'' \emph{\BIBforeignlanguage{en}{New J. Phys.}}, vol.~19, no.~2, p.
  023050, Feb. 2017.

\bibitem{Cong2021-bj}
I.~Cong, H.~Levine, A.~Keesling, D.~Bluvstein, S.-T. Wang, and M.~D. Lukin,
  ``{Hardware-Efficient}, {Fault-Tolerant} quantum computation with rydberg
  atoms,'' \emph{Physical Review X}, vol.~12, no.~2, p. 021049, 2022.

\bibitem{Jaksch2000-fh}
D.~Jaksch, J.~I. Cirac, P.~Zoller, S.~L. Rolston, R.~Cote, and M.~D. Lukin,
  ``\BIBforeignlanguage{en}{Fast quantum gates for neutral atoms},''
  \emph{\BIBforeignlanguage{en}{Phys. Rev. Lett.}}, vol.~85, no.~10, pp.
  2208--2211, Sep. 2000.

\bibitem{Endres2016-ez}
M.~Endres, H.~Bernien, A.~Keesling, H.~Levine, E.~R. Anschuetz, A.~Krajenbrink,
  C.~Senko, V.~Vuletic, M.~Greiner, and M.~D. Lukin,
  ``\BIBforeignlanguage{en}{Atom-by-atom assembly of defect-free
  one-dimensional cold atom arrays},'' \emph{\BIBforeignlanguage{en}{Science}},
  vol. 354, no. 6315, pp. 1024--1027, Nov. 2016.

\bibitem{Radnaev2024-lw}
A.~G. Radnaev, W.~C. Chung, D.~C. Cole, D.~Mason, T.~G. Ballance, M.~J.
  Bedalov, D.~A. Belknap, M.~R. Berman, M.~Blakely, I.~L. Bloomfield, P.~D.
  Buttler, C.~Campbell, A.~Chopinaud, E.~Copenhaver, M.~K. Dawes, S.~Y.
  Eubanks, A.~J. Friss, D.~M. Garcia, J.~Gilbert, M.~Gillette, P.~Goiporia,
  P.~Gokhale, J.~Goldwin, D.~Goodwin, T.~M. Graham, C.~J. Guttormsson, G.~T.
  Hickman, L.~Hurtley, M.~Iliev, E.~B. Jones, R.~A. Jones, K.~W. Kuper, T.~B.
  Lewis, M.~T. Lichtman, F.~Majdeteimouri, J.~J. Mason, J.~K. McMaster, J.~A.
  Miles, P.~T. Mitchell, J.~D. Murphree, N.~A. Neff-Mallon, T.~Oh, V.~Omole,
  C.~P. Simon, N.~Pederson, M.~A. Perlin, A.~Reiter, R.~Rines, P.~Romlow, A.~M.
  Scott, D.~Stiefvater, J.~R. Tanner, A.~K. Tucker, I.~V. Vinogradov, M.~L.
  Warter, M.~Yeo, M.~Saffman, and T.~W. Noel, ``A universal neutral-atom
  quantum computer with individual optical addressing and non-destructive
  readout,'' \emph{arXiv [quant-ph]}, Aug. 2024.

\bibitem{Pecorari2025-bl}
L.~Pecorari, S.~Jandura, G.~K. Brennen, and G.~Pupillo,
  ``\BIBforeignlanguage{en}{High-rate quantum {LDPC} codes for
  long-range-connected neutral atom registers},''
  \emph{\BIBforeignlanguage{en}{Nat. Commun.}}, vol.~16, no.~1, p. 1111, Jan.
  2025.

\bibitem{Grassl1996-ua}
M.~Grassl, T.~Beth, and T.~Pellizzari, ``Codes for the quantum erasure
  channel,'' \emph{Physical Review A}, vol.~56, no.~1, p.~33, 1997.

\bibitem{Dennis2001-jj}
E.~Dennis, A.~Kitaev, A.~Landahl, and J.~Preskill, ``Topological quantum
  memory,'' \emph{Journal of Mathematical Physics}, vol.~43, no.~9, pp.
  4452--4505, 2002.

\bibitem{Horsman2011-yv}
C.~Horsman, A.~G. Fowler, S.~Devitt, and R.~Van~Meter,
  ``\BIBforeignlanguage{en}{Surface code quantum computing by lattice
  surgery},'' \emph{\BIBforeignlanguage{en}{New J. Phys.}}, vol.~14, no.~12, p.
  123011, Dec. 2012.

\bibitem{Fowler2012-hs}
A.~G. Fowler, A.~C. Whiteside, A.~L. McInnes, and A.~Rabbani, ``Topological
  code autotune,'' \emph{Physical Review X}, vol.~2, no.~4, p. 041003, 2012.

\bibitem{Delfosse2017-be}
N.~Delfosse and G.~Zémor, ``\BIBforeignlanguage{en}{Linear-time maximum
  likelihood decoding of surface codes over the quantum erasure channel},''
  \emph{\BIBforeignlanguage{en}{Phys. Rev. Research}}, vol.~2, no.~3, Jul.
  2020.

\bibitem{IOlius2023-ty}
A.~deMarti iOlius, P.~Fuentes, R.~Or{\'u}s, P.~M. Crespo, and J.~E. Martinez,
  ``Decoding algorithms for surface codes,'' \emph{Quantum}, vol.~8, p. 1498,
  2024.

\bibitem{higgott2023sparse}
O.~Higgott and C.~Gidney, ``Sparse blossom: correcting a million errors per
  core second with minimum-weight matching,'' \emph{Quantum}, vol.~9, p. 1600,
  2025.

\bibitem{gidney2021stim}
\BIBentryALTinterwordspacing
C.~Gidney, ``Stim: a fast stabilizer circuit simulator,'' \emph{{Quantum}},
  vol.~5, p. 497, Jul. 2021. [Online]. Available:
  \url{https://doi.org/10.22331/q-2021-07-06-497}
\BIBentrySTDinterwordspacing

\bibitem{Kolmogorov2009-eb}
V.~Kolmogorov, ``Blossom v: a new implementation of a minimum cost perfect
  matching algorithm,'' \emph{Math. Program. Comput.}, vol.~1, no.~1, pp.
  43--67, Jul. 2009.

\bibitem{auger2017fault}
J.~M. Auger, H.~Anwar, M.~Gimeno-Segovia, T.~M. Stace, and D.~E. Browne,
  ``Fault-tolerance thresholds for the surface code with fabrication errors,''
  \emph{Physical Review A}, vol.~96, no.~4, p. 042316, 2017.

\bibitem{Whiteside2014-rc}
A.~C. Whiteside and A.~G. Fowler, ``Upper bound for loss in practical
  topological-cluster-state quantum computing,'' \emph{Physical Review A},
  vol.~90, no.~5, p. 052316, 2014.

\bibitem{Sahay2023-av}
K.~Sahay, J.~Jin, J.~Claes, J.~D. Thompson, and S.~Puri, ``High-threshold codes
  for neutral-atom qubits with biased erasure errors,'' \emph{Physical Review
  X}, vol.~13, no.~4, p. 041013, 2023.

\bibitem{McEwen2021-zn}
M.~McEwen, D.~Kafri, Z.~Chen, J.~Atalaya, K.~J. Satzinger, C.~Quintana, P.~V.
  Klimov, D.~Sank, C.~Gidney, A.~G. Fowler, F.~Arute, K.~Arya, B.~Buckley,
  B.~Burkett, N.~Bushnell, B.~Chiaro, R.~Collins, S.~Demura, A.~Dunsworth,
  C.~Erickson, B.~Foxen, M.~Giustina, T.~Huang, S.~Hong, E.~Jeffrey, S.~Kim,
  K.~Kechedzhi, F.~Kostritsa, P.~Laptev, A.~Megrant, X.~Mi, J.~Mutus,
  O.~Naaman, M.~Neeley, C.~Neill, M.~Niu, A.~Paler, N.~Redd, P.~Roushan, T.~C.
  White, J.~Yao, P.~Yeh, A.~Zalcman, Y.~Chen, V.~N. Smelyanskiy, J.~M.
  Martinis, H.~Neven, J.~Kelly, A.~N. Korotkov, A.~G. Petukhov, and R.~Barends,
  ``\BIBforeignlanguage{en}{Removing leakage-induced correlated errors in
  superconducting quantum error correction},''
  \emph{\BIBforeignlanguage{en}{Nat. Commun.}}, vol.~12, no.~1, p. 1761, Mar.
  2021.

\bibitem{Kozen1992}
\BIBentryALTinterwordspacing
D.~C. Kozen, \emph{Union-Find}.\hskip 1em plus 0.5em minus 0.4em\relax New
  York, NY: Springer New York, 1992, pp. 48--51. [Online]. Available:
  \url{https://doi.org/10.1007/978-1-4612-4400-4\_10}
\BIBentrySTDinterwordspacing

\bibitem{Fowler2013-sr}
A.~G. Fowler, ``Coping with qubit leakage in topological codes,'' \emph{Phys.
  Rev. A}, vol.~88, no.~4, p. 042308, Oct. 2013.

\bibitem{Fowler2012-hz}
A.~G. Fowler, M.~Mariantoni, J.~M. Martinis, and A.~N. Cleland, ``Surface
  codes: Towards practical large-scale quantum computation,'' \emph{Phys. Rev.
  A}, vol.~86, no.~3, p. 032324, Sep. 2012.

\bibitem{Norcia2023-zh}
M.~A. Norcia, W.~B. Cairncross, K.~Barnes, P.~Battaglino, A.~Brown, M.~O.
  Brown, K.~Cassella, C.-A. Chen, R.~Coxe, D.~Crow, J.~Epstein, C.~Griger,
  A.~M.~W. Jones, H.~Kim, J.~M. Kindem, J.~King, S.~S. Kondov, K.~Kotru,
  J.~Lauigan, M.~Li, M.~Lu, E.~Megidish, J.~Marjanovic, M.~McDonald,
  T.~Mittiga, J.~A. Muniz, S.~Narayanaswami, C.~Nishiguchi, R.~Notermans,
  T.~Paule, K.~A. Pawlak, L.~S. Peng, A.~Ryou, A.~Smull, D.~Stack, M.~Stone,
  A.~Sucich, M.~Urbanek, R.~J.~M. van~de Veerdonk, Z.~Vendeiro, T.~Wilkason,
  T.-Y. Wu, X.~Xie, X.~Zhang, and B.~J. Bloom,
  ``\BIBforeignlanguage{en}{Midcircuit qubit measurement and rearrangement in a
  {Yb171} atomic array},'' \emph{\BIBforeignlanguage{en}{Phys. Rev. X.}},
  vol.~13, no.~4, Nov. 2023.

\bibitem{PhysRevX.15.011009}
\BIBentryALTinterwordspacing
M.~Peper, Y.~Li, D.~Y. Knapp, M.~Bileska, S.~Ma, G.~Liu, P.~Peng, B.~Zhang,
  S.~P. Horvath, A.~P. Burgers, and J.~D. Thompson, ``Spectroscopy and modeling
  of $^{171}\mathrm{Yb}$ rydberg states for high-fidelity two-qubit gates,''
  \emph{Phys. Rev. X}, vol.~15, p. 011009, Jan 2025. [Online]. Available:
  \url{https://link.aps.org/doi/10.1103/PhysRevX.15.011009}
\BIBentrySTDinterwordspacing

\bibitem{Baker2021-oc}
J.~M. Baker, A.~Litteken, C.~Duckering, H.~Hoffmann, H.~Bernien, and F.~T.
  Chong, ``Exploiting long-distance interactions and tolerating atom loss in
  neutral atom quantum architectures,'' in \emph{2021 ACM/IEEE 48th Annual
  International Symposium on Computer Architecture (ISCA)}.\hskip 1em plus
  0.5em minus 0.4em\relax IEEE, Jun. 2021, pp. 818--831.

\bibitem{Manabe2023-va}
H.~Manabe, Y.~Suzuki, and A.~S. Darmawan, ``Efficient simulation of leakage
  errors in quantum error correcting codes using tensor network methods,''
  \emph{arXiv [quant-ph]}, Aug. 2023.

\bibitem{Google-Quantum-AI2023-yd}
{Google Quantum AI}, ``\BIBforeignlanguage{en}{Suppressing quantum errors by
  scaling a surface code logical qubit},''
  \emph{\BIBforeignlanguage{en}{Nature}}, vol. 614, no. 7949, pp. 676--681,
  Feb. 2023.

\bibitem{Chubb2021-in}
C.~T. Chubb and S.~T. Flammia, ``\BIBforeignlanguage{en}{Statistical mechanical
  models for quantum codes with correlated noise},''
  \emph{\BIBforeignlanguage{en}{Ann. Institut Henri Poincaré D}}, vol.~8,
  no.~2, pp. 269--321, May 2021.

\end{thebibliography}

\end{document}